\hoffset=-0.35cm
\voffset=1.0cm
\footline={\hfil}
\headline={\tenrm\hfil\folio}

\def\comp{{\rm C{\mkern -8.5mu}
           {{\vrule height 7.5pt width 0.85pt depth -0.83pt}}}{\mkern 8.5mu}}

\def\ket#1{| #1 \rangle}
\def\bra#1{\langle #1 |}
\def\scx#1#2{\langle #1\, |\, #2 \rangle} 
\def\bd#1{{\bf #1}}
\def\me#1#2#3{\langle #2\, |#1|\, #3\rangle}

\centerline{\bf{Resource Letter GPP-1: Geometric Phases in Physics}}

\vskip 0.45in
\centerline{Jeeva ~Anandan, ~Joy ~Christian, ~and ~Kazimir ~Wanelik}
\bigskip
\centerline{\it Department of Theoretical Physics, University of Oxford,
Oxford OX1 3NP, United Kingdom}
\vskip 0.45in
\baselineskip 0.5cm
\parindent 0.75cm
\centerline{ABSTRACT}

\bigskip
\bigskip

{\narrower{This Resource Letter provides a guide to the literature on the
geometric angles
and phases in classical and quantum physics. Journal articles and books are
cited for the following topics: anticipations of the geometric phase,
foundational derivations and formulations, books and review articles on the
subject, and theoretical and experimental elaborations and applications.}}

\bigskip
\bigskip
\bigskip

Suppose a system undergoes an evolution so that after some time it returns to
its original state. We shall call such an evolution a {\it cyclic} evolution.
If the system is classical, then it is impossible to say from its initial and
final states that it has undergone any evolution. However,
the wave function of a quantal system retains a memory of its motion in the
form of a geometric phase factor. This phase factor
can be measured by interfering the
wave function with another coherent wave function enabling one to discern
whether or not the system has undergone an evolution. Therefore
geometric phase factors are `signatures' of quantum motion. 
The adjective `geometric' emphasizes that such phase factors depend
only on the loop in the quantum-mechanical state space --- the set of rays
of the Hilbert space, sometimes called the projective Hilbert space. In
particular, geometric phases are independent of parameterization of the path
in the projective Hilbert space, and therefore of the speed at which
it has been traversed. 

As early as 1956 [1], in a classic paper on phase shifts in non-quantal
polarized light [2], S.~Pancharatnam anticipated the quantal geometric phases.
He was only twenty-two years of age at the time. He studied the problem of
determining the phase change undergone by polarized light after it has passed
through a sequence of polarizers such that its final polarization is the same
as its initial polarization. To describe how the phase of polarized light
changes under passage through a polarizer, Pancharatnam needed to define the
{\it phase difference} between two different polarization states. He reasoned
that the most natural way to accomplish this task is to ask what would happen
if two such states were brought to interfere with each other, and
accordingly he proposed the following definition: the polarization states
of any two monochromatic beams of light with the same momenta are {\it in phase}
if the superposition of the two has the maximum possible intensity. Let 
$\ket{A}$ and $\ket{B}$ represent the polarization state-vectors of photons in
the two beams. Since the intensity of their superposition is proportional to
$$
(\bra{A} + \bra{B})(\ket{A} + \ket{B})=2 + 
2\,|\scx{A}{B}|\cos\{{\rm ph}\scx{A}{B}\}\,,\eqno(1)
$$
according to his convention $\ket{A}$ and $\ket{B}$ are {\it in phase}
when their scalar product $\scx{A}{B}$ is real and positive, or equivalently,
when ${{\rm ph}\scx{A}{B}=0}$.
Incidentally, since orthogonal states do not interfere,
this convention breaks down for such states, and the phase difference between
them remains undefined. In the general case of {\it non-orthogonal} states,
it is natural to identify
the phase difference between $\ket{A}$ and $\ket{B}$ with the phase
${{\rm ph}\scx{A}{B}}$ of their scalar product.

Pancharatnam used this definition of the phase difference to analyze an
experiment involving a sequence of changes in polarization of a beam of
classical light by sending it through suitable polarizers. His
experiment consisted of three sequential changes in polarization, 
from $\ket{A}$ to $\ket{B}$ to $\ket{C}$ and back to a state $\ket{A'}$ of the
initial polarization. It is easy to show that in such a scheme
each successive state remains in phase with the previous one. Now, the label
$A$ used here to describe a state of a polarized wave of light represents a
set of values (the eigenvalues of a complete set of commuting
observables) required to specify this state uniquely. 
In Pancharatnam's experiment all but one of these values --- 
including the one that specifies the polarization --- were returned to their
original values, with the phase of polarization being the only exception.
Thus, Pancharatnam's evolution was not cyclic in the sense described above.
Indeed, in what follows, the classical phase difference he observed will be
shown to come from the quantum mechanical phase difference between the initial
and final one-photon states,
$$
\scx{\psi}{\psi'}={\rm e}^{-{{\rm i}\over 2}\alpha_{ABC}},\eqno(2)
$$
where $\alpha_{ABC}$ is the solid angle subtended by the geodesic
triangle $ABC$ on the Poincar\'e sphere (whose points, as is well-known,
represent all conceivable forms of polarization states). For simplicity, we
ignore the dynamical phase difference due to the fixed frequency of the
photon. Remarkably enough,
Pancharatnam not only anticipated the quantal geometric phases, but also
was able to corroborate his theory experimentally.

Another geometric phase given by a solid-angle formula analogous to (2) 
was put forward in 1984 by M.V.~Berry (who was unaware of Pancharatnam's work)
in a seminal paper on the quantum-mechanical adiabatic theorem [7]. He
investigated the nonrelativistic Schr\"odinger evolution 
$$
{\rm i}{{\rm d}\over{{\rm d}t}}
\ket{\psi(t)}=H(\bd{R}(t))\ket{\psi(t)}\eqno(3)
$$
of a quantal system in a slowly
changing environment described by a set of $N$ time-dependent parameters 
${\bd{R}(t)=(R_1(t), R_2(t),\dots,R_N(t))\,}$, with the initial state
$$
\ket{\psi(0)} = \ket{n; \bd{R}(0)}\eqno(4)
$$ 
being the stationary state given by the time-independent
Schr\"odinger equation
$$
H(\bd{R}(0))\ket{n; \bd{R}(0)} = 
E_n(\bd{R}(0))\ket{n;\bd{R}(0)}\,.\eqno(5)
$$
If $H(\bd{R}(t))$ is non-degenerate and slowly varying, then it is known that
the time-evolving Schr\"odinger state $\ket{\psi(t)}$ remains an eigenstate of
the instantaneous Hamiltonian $H(\bd{R}(t))$. More precisely,
$$
\ket{\psi(t)}=
{\rm e}^{-{\rm i}\int_0^t{\rm d}s\; E_n(\bd{R}(s))}
{\rm e}^{{\rm i}\, b\, [n; \bd{R}(t)]}\ket{E_n(\bd{R}(t))},\eqno(6)
$$
where
$$
b\, [n; \bd{R}(t)]=\int_0^t{\rm d}s\;  
\me{{\rm i}{{\rm d}\over{{\rm d}s}}}{n; \bd{R}(s)}{n; \bd{R}(s)}\,,\eqno(7)
$$
or, equivalently,
$$
b\, [n; \bd{R}(t)]=\int_{\bd{R}(0)}^{\bd{R}(t)}{\rm d}\bd{R'}\,\cdot\,
\me{{\rm i}\nabla_{\bd R'}}{n; \bd{R'}}{n; \bd{R'}}\,,\eqno(7')
$$
were ${\nabla_{\bd{R}}}$ is the gradient operator in the parameter space
${\bd{R}}$. This is, of course, just the time-honored adiabatic theorem.

Berry's investigations, however,
went beyond the usual formulation of the adiabatic theorem
captured in (6) and (7). He considered the case of an adiabatic transport
around a {\it closed} path
$$
\rho_t=\{\bd{R}(t)\,|\; \bd{R}(T)=\bd{R}(0)\,;\, 0<t<T\,\}\eqno(8)
$$
in the parameter
space, and made the crucial observation that in such an adiabatic setup the
phase factor ${e^{i\,b\, [n; \bd{R}(t)]}}$ is not
integrable; i.e., in general it cannot be written as a function of $\bd{R}$,
and in particular is not single-valued under continuation around the loop:
${e^{i\,b\, [n; \bd{R}(T)]}\ne e^{i\,b\, [n;\bd{R}(0)]}}$. 
Moreover, it is easy to see that $(7')$ can be reexpressed in the form 
$$
b\, [n; \rho]=\oint_{\rho}{\rm d}\bd{R}\,\cdot\,
\me{{\rm i}\nabla_{\bd{R}}}{n;\bd{R}}{n;\bd{R}}\,,\eqno(9)
$$ 
from which it is evident that ${b[n; \bd{R}(T)]}$, or the Berry phase as it is
now called, is independent of parameterization:
${b[n; \bd{R}(T)]=b[n; \rho]}$,
where $\rho$ denotes the unparameterized loop corresponding to $\rho_t$. In
particular, unlike the usual dynamical phase ${\{-\int {\rm d}s\,E_n\}\,}$,
the Berry phase ${b[n; \rho]}$
is independent of the rate at which the state of the
system traverses around ${\rho\,}$.

To illustrate his findings Berry analyzed the example of a spin-$s$
particle interacting with a magnetic field $\bd{B}$ through the
Hamiltonian 
$$
H(\bd{B})=\kappa\,\bd{B}{\cdot}\bd{S}\,,\eqno(10)
$$
where $\kappa$ is a constant involving the gyromagnetic ratio, and
$\bd{S}$ is the vector spin operator whose components have
$2s+1$ eigenvalues $n$ lying
between $-s$ and $+s$ with integer spacing. The eigenvalues
of $H(\bd{B})$ are, of course,
$$
E_n(\bd{B})=\kappa\,B\,n\;,\eqno(11)
$$
with ${B=|\bd{B}|}$. Now, if one identifies the components of the external
magnetic field $\bd{B}$ with the parameter space $\bd{R}$, then Berry's formula
is easily applicable to this case. In particular, (9) gives the geometric
phase change
of an eigenstate $\ket{n; \bd{B}(t)}$ of $H(\bd{B}(t))$ as $\bd{B}(t)$ 
is slowly transported --- and hence the spin is slowly precessed --- around a
loop $\gamma$ in the $\bd{B}$-space. Berry was able to show that, in that case,
$$
{\rm e}^{{\rm i}b\, [n; \bd{B}(T)]} = 
{\rm e}^{-{\rm i}n\alpha_{\gamma}},\eqno(12)
$$
where $\alpha_{\gamma}$ is the solid angle subtended by the loop $\gamma$ at
$\bd{B}=0$. In particular, when $s={1\over 2}$ and the initial state
is $\ket{{1\over 2}; \bd{B}(0)}$ (`spin up' along $\bd{B}$), then the
right-hand side of (12) takes the form of the right-hand side
of the observation
(2) of Pancharatnam --- namely, ${\rm e}^{-{{\rm i}\over 2}\alpha_{\gamma}}$.
To establish an analogy between (2) and (12) it suffices now to identify
$\ket{\psi(0)}=\ket{\psi}$ and $\ket{\psi(T)}=\ket{\psi'}$, and note that the
left-hand side of (12) can be rewritten, after the dynamical phase is
removed, in the form $\scx{\psi(0)}{\psi(T)}$.

A simple explanation of the beautiful result (12) was given in 1987
by J.~Anandan and L.~Stodolsky [36]. They considered a sphere whose points
represented the possible directions of the magnetic field ${\bf B}$. In the
above example of Berry, it is sufficient for the direction of ${\bf B}$ to
trace a closed curve ${\hat{\gamma}}$
on this sphere in order for each eigenstate to
acquire a geometric phase. In other words, it is not necessary for
${{\bf B}(t)}$ to form a closed curve; it is sufficient if merely the
directions of ${{\bf B}(0)}$ and ${{\bf B}(T)}$ coincide. Anandan and
Stodolsky then considered a Cartesian triad with its origin on
${\hat{\gamma}(t)}$ and its z-axis in the radial direction of the sphere
(the direction of ${{\bf B}(t)}$ and the spin axis). If the triad is moved
along ${\hat{\gamma}(t)}$ so that the x,y-axes are parallel-transported along
the surface of the sphere, then when the triad returns to the original point
${\hat{\gamma}(0)=\hat{\gamma}(T)}$, it will have rotated about its z-axis by
the solid angle $\alpha$ subtended by ${\hat{\gamma}}$ at the center of the
sphere. Now, relative to the triad, each eigenstate individually should
acquire only the usual dynamical phase factor because the triad has no angular
velocity about the spin axis. Consequently, the additional phase factor
acquired by
the eigenstate must be interpreted as the {\it geometric} phase factor due to
the rotation of the triad given by
${e^{i\alpha J_z}\ket{n}=e^{i\alpha n}\ket{n}}$,
where ${J_z}$ generates rotation about the z-axis of the triad.

For an arbitrary cyclic evolution in any Hilbert space, the above angle
$\alpha$ generalizes to a set of angles ${\alpha_1, \dots, \alpha_N}$. These
are the geometric quantum angles introduced by Anandan [14], which perhaps
provides the deepest approach so far to the geometric phase. The geometric
phases acquired by a complete set of orthogonal states
${\{\ket{n}\}}$ are now obtained by the action on each ${\ket{n}}$ by
${e^{i\sum_{k=1}^N\,\alpha_k\,J_k}}$, where the elements of the set ${\{J_k\}}$
commute among themselves. In the classical limit the geometric angles
${\{\alpha_k\}}$ reduce to the classical angles of J.H.~Hannay [10], while the
observables ${\{J_k\}}$ become the corresponding action variables that are in
involution with each other.

We now illustrate the usefulness of geometric angles by providing a
quantum-mechanical
explanation of the above-mentioned experiment of Pancharatnam. For
this purpose, we need to generalize his classical electromagnetic
polarized wave, with fixed momentum ${\bf p}$, passing through an
arbitrary number of polarizers, such that the
final polarization is the same as the initial polarization.
A classical electromagnetic wave is an approximation of a coherent
state in quantum electrodynamics. In the Coulomb gauge the quantized
vector-potential for the electromagnetic field may be written as
$$
{\bf A} = \sum_{\bf k}\sum_{\lambda=1}^2 [
a_{{\bf k},\lambda}\,e^{-i({\bf k}\cdot{\bf x}-\omega t)}\,+\,
a^{\dag}_{{\bf k},\lambda}\,e^{-i({\bf k}\cdot{\bf x}-\omega t)}]\,
{\bf e}_{{\bf k},\lambda}\;,\eqno(13)
$$
where ${\bf k}$ is the momentum vector, ${\omega = |{\bf k}|}$ is the
frequency, ${{\bf e}_{{\bf k},\lambda}}$
are real orthogonal polarization
vectors perpendicular to ${\bf k}$, and ${a_{{\bf k},\lambda}}$,
${a^{\dag}_{{\bf k},\lambda}}$ are the annihilation and creation operators for
the mode ${({\bf k},\lambda)}$. The electric and magnetic fields corresponding
to ${\bf A}$ are ${{\bf E} = -{{\partial{\bf A}}\over{\partial t}}}$ and
${{\bf B} = {\bf\nabla}\times{\bf A}}$, respectively. The coherent state
corresponding to the electromagnetic wave considered by Pancharatnam is then
$$
\ket{z_1,\,z_2,\,{\bf p}}\;=\; e^{-{1\over 2}(|z_1|^2+|z_2|^2)}\,
\exp({z_1a^{\dag}_{{\bf p},1}}\,+\,{z_2a^{\dag}_{{\bf p},2}})\,
\ket{0}\;,\eqno(14)
$$
which is an eigenstate of ${a_{{\bf p},\lambda}}$ with eigenvalues
${z_{\lambda}}$. Therefore,
$$
\bra{z_1,\,z_2,\,{\bf p}}{\bf A}\ket{z_1,\,z_2,\,{\bf p}} \,=\, 
2\,\{|z_1|\cos{({\bf p}\cdot{\bf x}-\omega t+\theta_1)}\,{\bf e}_{{\bf p},1}\,+
\,|z_2|\cos{({\bf p}\cdot{\bf x}-\omega t+\theta_2)}\,{\bf e}_{{\bf p},2}\}
\;,\eqno(15)
$$
where ${\theta_1}$ and ${\theta_2}$ are the phases of ${z_1}$ and ${z_2}$
respectively.
It follows that ${|z_1|\omega}$ and ${|z_2|\omega}$ are the amplitudes of the
electric field ${\bf E}$ in the directions of ${{\bf e}_{{\bf p},1}}$
and ${{\bf e}_{{\bf p},2}}\,$, respectively.

We may represent the polarization state of a one-photon state
${({z_1a^{\dag}_{{\bf p},1}}\,+\,{z_2a^{\dag}_{{\bf p},2}})\,\ket{0}}$ as a
two-component spinor
${\left(\matrix{z_1\cr z_2}\right)}$ in a two dimensional vector space with
the usual inner product, which makes it a Hilbert space. The corresponding
projective Hilbert space is the Poincar\'e sphere. As each photon corresponding
to the mode ${({\bf p}, \mu)}$ passes through the polarizer it undergoes a
transition to a state ${({\bf p}, \mu')}$. The new state is obtained by simply
projecting the old state onto the state that passes through the polarizer. It
can be shown that this corresponds to parallel-transporting the old
state-vector
along the shorter geodesic joining the two points on the Poincar\'e
sphere representing the two polarization states [43]. Therefore, using
arguments similar to those used by Anandan and Stodolsky [36], the final state
obtained after a sequence of such polarization changes that return the photons
to their initial polarization state is given by the action of the operator
${e^{i\alpha J}}$ on the initial photon state. Here ${\alpha}$ is the
solid angle subtended by the geodesic polygon defined by the sequence of the
polarization states on the Poincar\'e sphere, and ${J={N\over 2}}$, $N$ being
the number operator for the initial and final mode ${({\bf p}, \mu)}$.
As a result, the final state of the electromagnetic field is
$$
e^{i\alpha J}\,\ket{z_1,\,z_2,\,{\bf p}}\;=\;
\ket{z_1e^{i{{\alpha}\over 2}},\,z_2e^{i{{\alpha}\over 2}},\,{\bf p}}
\;.\eqno(16)
$$
Finally, in this resultant state we have
$$
\bra{z_1,\,z_2,\,{\bf p}}{\bf A}\ket{z_1,\,z_2,\,{\bf p}} \,=\, 2\,
\{|z_1|\cos{({\bf p}\cdot{\bf x}-\omega t+\theta_1+{{\alpha}\over 2})}\,
{\bf e}_{{\bf p},1}\,+\,|z_2|\cos{({\bf p}\cdot{\bf x}-\omega t+\theta_2
+{{\alpha}\over 2})}\,{\bf e}_{{\bf p},2}\}\;\eqno(17)
$$
Comparison of this expression with equation (15)
shows that ${{\alpha}\over 2}$
is the phase Pancharatnam observed in his classical experiment.
A similar explanation can be given to the
experiment of Tomita and Chiao [117], except in their case we would have
${J=N}$ for the photon since it is a spin-1 particle.

Having obtained these results using the operator ${\exp(i\alpha J)}$, which
depends on the geometric angle ${\alpha}$, we may generalize them to an
arbitrary superposition of number eigenstates with the same polarization. In
general, such a state would not be a coherent state and cannot therefore be
represented by a classical electromagnetic wave. Nevertheless, the geometric
part of the evolution may be obtained by taking the expectation value of
${\exp(i\alpha J)}$ with respect to the initial state.

The above mentioned geometric treatment of Berry's phase by Anandan and
Stodolsky suggest that the geometric phase is associated with the motion of a
quantum system and not with the particular Hamiltonian used to achieve this
motion. This is the basic idea used by Aharonov and Anandan [11] in obtaining
a geometric phase, which, since it is associated with the motion of the
quantum-mechanical state itself, does not require an adiabatically
varying Hamiltonian (environment). However, if an adiabatically varying
Hamiltonian is used to implement this motion, then this geometric phase is the
same as Berry's phase. They defined the evolution of a normalized state
$\ket{\psi(t)}$ to be cyclic in the interval $[0,T]$ if and only if
$$
\ket{\psi(T)}={\rm e}^{{\rm i}\phi(0,T)}\ket{\psi(0)},\eqno(18)
$$
where $\phi(0,T)$ is a real number. Equivalently, this can be reexpressed with
the help of the unitary time evolution operator $U(0,t)$ in the form
$$
U(0,T)\ket{\psi(0)}={\rm e}^{{\rm i}\phi(0,T)}\ket{\psi(0)}.\eqno(19)
$$
It follows from this equation that for an initial state to evolve
cyclicly in the interval $[0,T]$ under the time-evolution operator 
$U(0,t)$, it is necessary and sufficient for it to be an eigenstate of the
operator $U(0,T)$. Incidentally, this assures the existence of cyclic
evolutions as defined above at least in the finite-dimensional case. According
to Aharonov and Anandan, the geometric contribution to $\phi(0,T)$,
denoted by $\beta$, is
$$
{\rm e}^{{\rm i}\beta}={\rm e}^{{\rm i}\phi(0,T)+{\rm i}
\int_0^T{\rm d}s\, \me{{\rm i}{{\rm d}\over{\rm d}s}}
{\psi(s)}{\psi(s)}}\eqno(20)
$$
or equivalently,
$$
{\rm e}^{{\rm i}\beta}=\scx{\psi(0)}{\psi(T)}\, {\rm e}^{{\rm i}
\int_0^T{\rm d}s\, \me{{\rm i}{{\rm d}\over{\rm d}s}}
{\psi(s)}{\psi(s)}}\,,\eqno(20')
$$
which reduces to the Berry's phase factor in the adiabatic limit [11, 67].
What is more, just as
Berry's phase, it is independent of the choice of parameterization or the
speed at which the path $\ket{\psi(t)}$ is traversed. More significantly,
Aharonov and Anandan demonstrated that $\beta$ is projective-geometric in
nature; i.e., it is the same for all paths $\ket{\phi(t)}$
that project to the same path in the projective Hilbert space. In other words,
it is the same for any two motions ${\ket{\phi(t)}}$ and ${\ket{\psi(t)}}$ such
that ${\rm p}_{\phi(t)}={\rm p}_{\psi(t)}$, where ${\rm p}_{\alpha}$ denotes
a ray corresponding to a vector $\ket{\alpha}$, namely,
$$
{\rm p}_{\alpha}=\{\ket{\beta}\,|\; \ket{\beta}=z\ket{\alpha};\,
z\in\comp\}.
\eqno(21)
$$
The above two properties imply that $\beta=\beta\, [{\rm p}_{\psi}]$, and
suggest that $\beta$ may have a geometric interpretation in terms of
paths in the projective Hilbert space. Indeed, it may be geometrically
understood as the {\it anholonomy} with respect to the natural connection on
the projective Hilbert space.
This interpretation generalizes an earlier differential-geometric
interpretation of the Berry phase given by B.~Simon in 1983 [8].

Berry's 1984 paper was concerned with nondegenerate states undergoing
adiabatic evolution. In the same year F.~Wilczek and A.~Zee reported on
how the theory can be generalized to include the adiabatic evolution
of {\it degenerate} quantum states [9].
They showed that in the case of a $d$-fold
degeneracy Berry's phase factor of the nondegenerate case, ${\rm
e}^{{\rm i}b\, [n; \rho]}$, is generalized to a $d\times d$ unitary
matrix, which is now called the non-Abelian Berry phase or the Wilczek-Zee
phase. More precisely, if the initial
state is one of the eigenstates belonging to an orthonormal set of eigenstates
of $H(\bd{R}(0))$ with a $d$-fold degenerate eigenvalue
$E_n(\bd{R}(0))$, i.e.
$$
H(\bd{R}(0))\ket{l; \bd{R}(0)}=E_n(\bd{R}(0))
\ket{l; \bd{R}(0)}\eqno(22)
$$ 
with $l=1,2,\dots,d$, then
$$
\ket{\psi(t)}={\rm e}^{-{\rm i}\int_0^t{\rm d}s\; E_n(\bd{R}(s))}
\sum_{l'=1}^{d}D_{l'l}(\bd{R}(t))\ket{l'; \bd{R}(t)}.\eqno(23)
$$
Here, the matrix $D$ is a path-ordered exponential integral 
$$
D\, [\bd{R}(t)]=
{\cal P}\! \exp\left\{{\rm i}\int_{\bd{R}(0)}^{\bd{R}(t)}
{\rm d}\bd{R'}\;  A(\bd{R'})\right\},\eqno(24)
$$ 
with
$$
A_{l'l}(\bd{R}(t))
=\me{{\rm i}\nabla_{\bd{R}}}{l'; \bd{R}}{l; \bd{R}}\,,\eqno(25)
$$
and ${\cal P}$ represents the path-ordering.
The non-Abelian phase factor $D\, [l; \bd{R}(T)]$ is a unitary matrix,
and may be denoted by $D\, [l; \rho]$ because it too is independent of
parameterization or the speed with which a particular path is traversed
and is therefore geometric.

Berry's 1984 paper, and the other reports discussed above,
were followed by a great number of papers on the subject. They can be
divided into two broad groups. The first contains
contributions that reformulate or generalize Berry's findings,
while the second contains papers in which geometric phases are
identified or measured in a great number of apparently disparate
physical phenomena, or in which attempts are made to use these phases
to explain unresolved physical questions. 
The bibliography that follows
is selective and by no means exhaustive since
there are hundreds of research papers on the subject.

We conclude by remarking that, there are at least four different
reasons for the phenomenal success of the concepts related to geometric phases.
First, these concepts are exceptionally clear and have a very
elegant geometric interpretation in terms of anholonomies and connections
(gauge fields).
Second, geometric phases have a certain unifying character that enables
one to relate many apparently disparate phenomena.
Third, these phases can be observed and, indeed, various predictions of the
geometric phases have been amply corroborated.
Finally, and perhaps most importantly, these concepts reassert the importance
and fruitfulness of geometric ideas in physical theories.

The work of one of the authors (J. A.) was partially supported by NSF grant no.
PHY-9307708 and ONR grant no. R \& T 3124141. 

\bigskip

\item{{\bf I.}} {\bf ANTICIPATIONS OF THE GEOMETRIC PHASE}

\item{1.} Actually, there are some Russian theoretical papers, collected in the
reference [18] below, which anticipate the geometric phase as early as the
1940s.

\item{2.}``Generalized theory of interference, and its applications,'' S.
Pancharatnam, Proc. Indian Acad. Sci. A {\bf 44}, 247-262 (1956). See also
reference [43] below for the formulation of Pancharatnam's phase in quantum
theoretical language. (E)

\item{3.}``Significance of electromagnetic potentials in the quantum theory,''
Y. Aharonov and D. Bohm, Phys. Rev. {\bf 115} 485-491 (1959). (I)

\item{4.}``Intersection of potential energy surfaces in polyatomic molecules,''
G. Herzberg and H.C. Longuet-Higgins, Disc. Farad. Soc. {\bf 35}, 77-82 (1963).
(I)

\item{5.}``Spin-orbit coupling and the intersection of potential energy
surfaces in polyatomic molecules,'' A.J. Stone, Proc. R. Soc. Lond. A
{\bf 351}, 141-150 (1976). (I)

\item{6.}``On the determination of Born-Oppenheimer nuclear motion wave
functions including complications due to conical intersections and identical
nuclei,'' C.A. Mead, J. Chem. Phys. {\bf 70}, 2284-2296 (1979). (I)

\bigskip

\item{{\bf II.}} {\bf FOUNDATIONAL DERIVATIONS AND FORMULATIONS}

\item{7.} ``Quantum phase factors accompanying adiabatic changes,'' M.V. Berry,
Proc. R. Soc. Lond. A {\bf 392}, 45-57 (1984). (I)

\item{8.} ``Holonomy, the quantum adiabatic theorem, and Berry's phase,'' B.
Simon, Phys. Rev. Lett. {\bf 51}, 2167-2170 (1983). (A)

\item{9.} ``Appearance of gauge structure in simple dynamical systems,'' F.
Wilczek and A. Zee, Phys. Rev. Lett. {\bf 52}, 2111-2114 (1984). (I)

\item{10.} ``Angle Variable Holonomy in Adiabatic Excursion of an Integrable
Hamiltonian,'' J.H. Hannay, J. Phys. A {\bf 18}, 221-230 (1985). (I)

\item{11.} ``Phase change during a cyclic quantum evolution,'' Y. Aharonov and
J. Anandan, Phys. Rev. Lett. {\bf 58}, 1593-1596 (1987). (I)

\item{12.} ``General setting for Berry's phase,'' J. Samuel and R. Bhandari,
Phys. Rev. Lett. {\bf 60}, 2339-2342 (1988). (A)

\item{13.} ``Non-adiabatic non-abelian geometric phase,'' J. Anandan,
Phys. Lett. A {\bf 133}, 171-175 (1988). (I)

\item{14.} ``Geometric angles in quantum and classical physics,'' J. Anandan,
Phys. Lett. A {\bf 129}, 201-207 (1988). (I)

\item{15.} ``Geometrical Phases from Global Gauge Invariance of Non Linear
Classical Field Theories,'' J.C. Garrison and R.Y. Chiao, Phys. Rev. Lett.
{\bf 60}, 165-168 (1988). (I)

\item{16>} ``Comment on Geometric Phases for Classical Field Theories,''
J. Anandan,  Phys. Rev. Lett. {\bf 60}, 2555 (1988). (I)

\bigskip

\item{{\bf III.}} {\bf BOOKS AND REVIEW ARTICLES}

\item{17.} {\bf Geometric Phases in Physics}, edited by A. Shapere and F.
Wilczek
(World Scientific, Singapore, 1989). This book contains original reprints
of many pioneering articles on the subject with some introductory comments on
each sub-topic, from elementary to advanced. (E, I, A)

\item{18.} {\bf Topological Phases in Quantum Theory}, edited by B. Markovski
and S.I. Vinitsky (World Scientific, Singapore, 1989).
This book traces anticipations of the
geometric phase emphasizing Russian contributions to the subject and contains
many less well-known reprints. (E, I, A)

\item{19.} ``Berry's topological phase in quantum-mechanics and quantum-field
theory,'' I.J.R. Aitchison, Physica Scripta T {\bf 23}, 12-20 (1988). (I) 

\item{20.} ``Adiabatic quantum transport in multiply connected systems,'' J.E.
Avron, B. Zur, and A. Raveh, Rev. Mod. Phys. {\bf 60}, 873-915 (1988). (A)

\item{21.} ``Berry phase.'' J.W. Zwanziger, M. Koenig, and A. Pines, Annu. Rev.
Phys. Chem. {\bf 41}, 601-646 (1990). This review contains an extensive list of
articles up to 1990. (I)

\item{22.} ``Anticipations of the geometric phase,'' M.V. Berry, Physics Today
{\bf 43}, 34-40 (1990). (E)

\item{23.} ``Topological phases in quantum-mechanics and polarization optics,''
S.I Vinitskii, V.L. Derbov, V.M. Dubovik, B.L. Markovski, and Y.P.
Stepanovskii, Uspekhi (Sov. Phys.) {\bf 33}, 403-428 (1990). (I)

\item{24.} ``The geometric phase,'' J. Anandan, Nature {\bf 360}, 307-313
(1992). (E)

\item{25.} ``The geometric phase in molecular-systems,'' C.A. Mead, Rev. Mod.
Phys. {\bf 64}, 51-85 (1992). (I)

\item{26.} {\bf Quantum Mechanics}, A. B\"ohm
(Springer-Verlag, New York, 1993).
The last two chapters of this book contains a detailed textbook introduction to
the subject. (E)

\item{27.} ``Pancharatnam memorial issue,'' Guest Editors: S. Ramaseshan and
R. Nityananda, Current Science, India, {\bf 67}, 217-294 (1994). This special
issue contains many original contributions in addition to some review articles.
(E)

\bigskip

\item{{\bf IV.}} {\bf ELABORATIONS AND APPLICATIONS}

\item{{\bf A.}} {\bf Theoretical articles}

\item{28.} ``Fractional Statistics and the Quantum Hall-Effect,'' D. Arovas,
J.R. Schrieffer, and F. Wilczek, Phys. Rev. Lett. {\bf 53}, 722-723 (1984). (A)

\item{29.} ``Hamiltonian Interpretation of Anomalies,'' P. Nelson and L.
Alvarez-Gaum\'e, Comm. Math. Phys. {\bf 99}, 103-114 (1985). (A)

\item{30.} ``Classical Adiabatic Angles and Quantal Adiabatic Phase,''
M.V. Berry, J. Phys. A {\bf 18}, 15-27 (1985). (I)

\item{31.} ``Semiclassical quantization with a quantum adiabatic phase,'' H.
Kuratsuji and S. Iida, Phys. Lett. A {\bf 111}, 220-222 (1985). (I)

\item{32.} ``Classical and quantum adiabatic invariants,'' E. Gozzi, Phys.
Lett. B {\bf 165}, 351-354 (1985). (I)

\item{33.} ``Quantum holonomy and the chiral gauge anomaly,''
A.J. Niemi and  G.W. Semenoff, Phys. Rev. Lett. {\bf 55}, 927-930 (1985). (A)

\item{34.} ``Effective action for adiabatic process -- dynamical meaning of
Berry and Simon phase,'' H. Kuratsuji and S. Iida,
Prog. Theo. Phys. {\bf 74}, 439-445 (1985). (I)

\item{35.} ``The interference of polarized light as an early example of Berry's
phase,'' S. Ramaseshan and R. Nityananda, Current Science, India, {\bf 55},
1225-1226 (1986). (E)

\item{36.} ``Some geometrical considerations of Berry phase,'' J. Anandan and
L. Stodolsky, Phys. Rev. D {\bf 35}, 2597-2600 (1987). (I)

\item{37.} ``Classical adiabatic holonomy and its canonical structure,'' E.
Gozzi and W.D. Thacker, Phys. Rev. D {\bf 35}, 2398-2406 (1987). (I)

\item{38.} ``Berry phases, magnetic monopoles, and Wess-Zumino terms or how the
skyrmion got its spin,'' I.J.R. Aitchison, Acta Physica Polonica B {\bf 18},
207-235 (1987). (A)

\item{39.} ``Non-abelian Berry phase effects and optical-pumping of atoms,'' J.
Segert, Ann. Phys. {\bf 179}, 294-312 (1987). (I)

\item{40.} ``Non-abelian Berry phase, accidental degeneracy, and
angular-momentum,'' J. Segert, J. Math. Phys. {\bf 28}, 2102-2114 (1987). (I)

\item{41.} ``A topological investigation of the quantum adiabatic phase,'' E.
Kiritsis, Comm. Math. Phys. {\bf 111}, 417-437 (1987). (A)

\item{42.} ``Direct calculations of the  Berry phase for spins and
helicities,'' T.F. Jordan, J. Math. Phys. {\bf 28}, 1759-1760 (1987). (E)

\item{43.} ``The adiabatic phase and Pancharatnam phase for polarized-light,''
M.V. Berry, J. Mod. Optics {\bf 34}, 1401-1407 (1987). (E)

\item{44.} ``Cyclic geometrical quantum phases -- group-theory derivation and
manifestations in atomic physics,'' C. Bouchiat, J. de Physique {\bf 48},
1401-1406 (1987). (I)

\item{45.} ``Interpreting the anholonomy of coiled light,'' M.V. Berry, Nature
{\bf 326}, 277-278 (1987). (I)

\item{46.} ``Quantum phase corrections from adiabatic iteration,'' M.V. Berry,
Proc. R. Soc. Lond. A {\bf 414}, 31-46 (1987). (I)

\item{47.} ``Appearance of gauge potentials in atomic collision physics,'' B.
Zygelman, Phys. Lett. A {\bf 125}, 476-481 (1987). (I)

\item{48.} ``Molecular Kramers degeneracy and non-abelian adiabatic
phase-factors,'' C.A. Mead, Phys. Rev. Lett. {\bf 59}, 161-164 (1987). (I)

\item{49.} ``Berry geometrical phase and the sequence of states in the
Jahn-Teller effect,'' F.S. Ham, Phys. Rev. Lett. {\bf 58}, 725-728 (1987). (I)

\item{50.} ``Induced gauge-fields in a nongauged quantum system,'' H.Z. Li,
Phys. Rev. Lett. {\bf 58}, 539-542 (1987). (I)

\item{51.} ``Geometrical description of Berry's phase,'' D.N. Page, Phys. Rev.
A {\bf 36}, 3479-3481 (1987). (A)

\item{52.} ``Geometric quantum phase and angles,'' J. Anandan and Y. Aharonov,
Phys. Rev. D {\bf 38}, 1863-1870 (1988). (I)

\item{53.} ``Berry phase, locally inertial frames, and classical analogs,''
M. Kugler and S. Shtrikman, Phys. Rev. D {\bf 37}, 934-937 (1988). (E)

\item{54.} ``Berry phase and unitary transformations,'' T.F. Jordan, J. Math.
Phys. {\bf 29}, 2042-2052 (1988). (I)

\item{55.} ``Non-integrable quantum phase in the evolution of a spin-1
system,''
C. Bouchiat and G.W. Gibbons, J. de Physique {\bf 49}, 187-199 (1988). (I)

\item{56..} ``Complex geometrical phases for dissipative systems,'' J.C.
Garrison and E.M. Wright, Phys. Lett. A {\bf 128}, 177-181 (1988). (I)

\item{57.} ``Non-abelian gauge structure in nuclear-quadrupole resonance,'' A.
Zee, Phys. Rev. A {\bf 38}, 1-6 (1988). (I)

\item{58.} ``Effective action for a nonadiabatic quantum process,'' A. Bulgac,
Phys. Rev. A {\bf 37}, 4084-4089 (1988). (I)

\item{59.} ``Cyclic evolution in quantum-mechanics and the phases of
Bohr-Sommerfeld and Maslov,'' R.G. Littlejohn, Phys. Rev. Lett. {\bf 61},
2159-2162 (1988). (I)

\item{60.} ``Geometric canonical phase-factors and path-integrals,'' H.
Kuratsuji, Phys. Rev. Lett. {\bf 61}, 1687-1690 (1988). (I)

\item{61.} ``Classical non-adiabatic angles,'' M.V. Berry and J.H. Hannay, J.
Phys. A {\bf 21}, L325-L331 (1988). (I)

\item{62.} ``The Berry phase and the Hannay angle,'' G. Ghosh and B. Dutta-Roy,
Phys. Rev. D {\bf 37}, 1709-1711 (1988). (I)

\item{63.} ``Non-abelian geometric phase from incomplete quantum
measurements,''
J. Anandan and A. Pines, Phys. Lett. A {\bf 141}, 335-339 (1989). (I)

\item{64.} ``Non local aspects of quantum phases,'' J. Anandan, Ann.
Inst. Henri Poincar\'e {\bf 49}, 271-286 (1988). (I)

\item{65.} ``On removing Berry phase,'' G. Giavarini, E. Gozzi, D. Rohrlich,
and W.D. Thacker, Phys. Lett. A {\bf 138}, 235-241 (1989). (I)

\item{66.} ``Berry phase and Fermi-Walker parallel transport,'' R. Dandoloff,
Phys. Lett. A {\bf 139}, 19-20 (1989). (I)

\item{67.} ``The Berry phase as an appropriate correspondence limit of the
Aharonov-Anandan phase in a simple model,'' J. Christian and A. Shimony,
in {\bf Quantum Coherence}, edited by J. Anandan (World Scientific,
Singapore, 1990) pp 121-135. (E)

\item{68.} ``Berry phase, interference of light-beams, and the Hannay angle,''
G.S. Agarwal and R. Simon, Phys. Rev. A {\bf 42}, 6924-6927 (1990). (E)

\item{69.} ``Connections of Berry and Hannay type for moving Lagrangian
submanifolds,''
A. Weinstein, Advances in Maths. {\bf 82}, 133-159 (1990). (A)

\item{70.} ``On quantum holonomy for mixed states,'' L. Dabrowski and H.
Grosse, Lett. Maths. Phys. {\bf 19}, 205-210 (1990). (I)

\item{71.} ``Geometric phase in neutron interferometry,'' A.G. Wagh and V.C.
Rakhecha, Phys. Lett. A {\bf 148}, 17-19 (1990). (I)

\item{72.} ``Geometric phase for cyclic motions and the quantum state-space
metric,'' J. Anandan, Phys. Lett. A {\bf 147}, 3-8 (1990). (I)

\item{73.} ``Non-abelian geometric phase and long-range atomic forces,'' B.
Zygelman, Phys. Rev. Lett. {\bf 64}, 256-259 (1990). (I)

\item{74.} ``A geometric approach to quantum mechanics,'' J. Anandan, Found.
Phys. {\bf 21}, 1265-1284 (1991). (A)

\item{75.} ``How much does the rigid body rotate -- a Berry phase from the
18th-century,'' R. Montgomery, Am. J. Phys. {\bf 59}, 394-398 (1991). (E)

\item{76.} ``A gauge field governing parallel transport along mixed states,''
A. Uhlmann, Lett. Maths. Phys. {\bf 21}, 229-236 (1991). (A)

\item{77.} ``Born-Oppenheimer revisited,'' Y. Aharonov, E. Benreuven, S.
Popescu, and D. Rohrlich, Nuclear Physics B {\bf 350}, 818-830 (1991). (I)

\item{78.} ``Budden and Smith additional memory and the geometric phase,''
M.V. Berry, Proc. R. Soc. Lond. A {\bf 431}, 531-537 (1990). (I)

\item{79.} ``On the real and complex geometric phases,'' I.J.R. Aitchison and
K. Wanelik, Proc. R. Soc. Lond. A {\bf 439}, 25-34 (1992). (I)

\item{80.} ``A group theoretical treatment of the geometric phase,'' E.C.G.
Sudarshan, J. Anandan, and T.R. Govindarajan, Phys. Lett. A {\bf 164}, 133-137
(1992). (A)

\item{81.} ``Origin of the geometric forces accompanying Berry's geometric
potentials,'' Y. Aharonov and A. Stern, Phys. Rev. Lett. {\bf 69}, 3593-3597
(1992). (A)

\item{82.} ``Geometric phases and symmetries in dissipative systems,'' A.S.
Landsberg, Phys. Rev. Lett. {\bf 69}, 865-868 (1992). (A)

\item{83.} ``Berry phase, motive forces, and mesoscopic conductivity,'' A.
Stern, Phys. Rev. Lett. {\bf 68}, 1022-1025 (1992). (A)

\item{84.} ``Geometric phase, geometric distance, and length of the curve in
quantum evolution,'' A.K. Pati, J. Phys. A {\bf 25}, L1001-L1008 (1992). (I)

\item{85.} ``The geometric phase for chaotic systems,'' J.M. Robbins and M.V.
Berry, Proc. R. Soc. Lond. A {\bf 436}, 631-661 (1992). (A)

\item{86.} ``Berry phase and Euclidean path integral,'' T. Kashiwa, S. Nima,
and S. Sakoda, Ann. Phys. {\bf 220}, 248-273 (1992). (A)

\item{87.} ``Cyclic states, Berry phases and the Schr\"odinger operator,'' A.N.
Seleznyova, J. Phys. A {\bf 26}, 981-1000 (1993). (A)

\item{88.} ``Connection between solitons and geometric phases,'' R.
Balakrishnan, Phys. Lett. A {\bf 180}, 239-243 (1993). (A)

\item{89.} ``Geometric phase for the relativistic Klein-Gordon equation,'' J.
Anandan and P.O. Mazur, Phys. Lett. A {\bf 173}, 116-120 (1993). (A)

\item{90.} ``Nonabelian Berry phases in Baryons,'' H.K. Lee, M.A. Nowak, M.
Rho, and I. Zahed, Ann. Phys. {\bf 227}, 175-205 (1993). (A)

\item{91.} ``Interplay of Aharonov-Bohm and Berry phases,'' B. Reznik and Y.
Aharonov, Phys. Lett. B {\bf 315}, 386-391 (1993). (A)

\item{92.} ``Nonlinearity of Pancharatnam's topological phase,'' H. Schmitzer,
S. Klein, and W. Dultz, Phys. Rev. Lett. {\bf 71}, 1530-1533 (1993). (A)

\item{93.} ``Berry phase and the magnus force for a vortex line in a
superconductor,'' P. Ao and D.J. Thouless, Phys. Rev. Lett. {\bf 70},
2158-2161 (1993). (A)

\item{94.} ``Quantum kinematical approach to the geometric phase,'' N. Mukunda
and R. Simon, Ann. Phys. {\bf 228}, 205-268 (1993). (I)

\item{95.} ``Geometric phase in vacuum instability -- applications in
quantum cosmology,'' D.P. Datta, Phys. Rev. D {\bf 48}, 5746-5750 (1993).
(A)

\item{96.} ``On geometric phases and dynamical invariants,'' D.B. Monteoliva,
H.J. Korsch, and J.A. Nunez, J. Phys. A {\bf 27}, 6897-6906 (1994). (I)

\item{97.} ``Symplectic structure for the non-abelian geometric phase,'' D.
Chruscinski, Phys. Lett. A {\bf 186}, 1-4 (1994). (A)

\item{98.} ``Spin-orbit interaction and Aharonov-Anandan phase in mesoscopic
rings,'' T.Z. Qian and Z.B. Su, Phys. Rev. Lett. {\bf 72}, 2311-2315 (1994).
(A)

\item{99.} ``Adiabatic approximation and Berry's phase in the Heisenberg
picture,'' Y. Brihaye and P. Kosinski, Phys. Lett. A {\bf 195}, 296-300
(1994). (I)

\item{100.} ``Topological interpretations of quantum Hall conductance,'' D.J.
Thouless, J. Math. Phys. {\bf 35}, 5362-5372 (1994). (A)

\item{101.} ``Geometric-phase effects in laser dynamics,'' V.Y. Toronov and V.
Derbov, Phys. Rev. A {\bf 50}, 878-881 (1994). (A)

\item{102.} ``Aharonov-Bohm and Berry phases for a quantum cloud of charge,''
Y. Aharonov, S. Coleman, A.S. Goldhaber, S. Nussinov, S. Popescu, B. Reznik,
D. Rohrlich, and L. Vaidman, Phys. Rev. Lett. {\bf 73}, 918-921 (1994). (I)

\item{103.} ``S-matrix as geometric phase factor,'' R.G. Newton, Phys. Rev.
Lett. {\bf 72}, 954-956 (1994). (A)

\item{104.}``Macroscopic polarization in crystalline dielectrics --
the geometric
phase approach,'' R. Resta, Rev. Mod. Phys. {\bf 66}, 899-915 (1994). (A)

\item{105.} ``Geometric phases and Mielnik's evolution loops,'' D.J. Fernandez,
Int. J. Theo. Phys. {\bf 33}, 2037-2047 (1994). (A)

\item{106.}``Geometric phase of polarized hydrogenlike atoms in an external
magnetic-field,'' Z. Tang and D. Finkelstein, Phys. Rev. Lett. {\bf 74},
3134-3137 (1995). (A)

\item{107.}``The geometric vector potential in molecular-systems with
arbitrarily many identical nuclei,''B.Kendrick and C.A. Mead, J. Chem.
Phys. {\bf 102}, 4160-4168 (1995). (A)

\item{108.}``Geometric phase effects for wave-packet revivals,'' C. Jarzynski,
Phys. Rev. Lett. {\bf 74}, 1264-1267 (1995). (A)

\item{109.}``New derivation of the geometric phase,'' A.K. Pati, 
Phys. Lett. A {\bf 202}, 40-45 (1995). (I)

\item{110.} ``Gravity and geometric phases,'' A. Corichi and M. Pierri,
Phys. Rev. D {\bf 51}, 5870-5875 (1995). (A)

\item{111.} ``Nature of geometric (Berry) potentials,'' A. Krakovsky and J.L. 
Birman, Phys. Rev. A {\bf 51}, 50-53 (1995). (I)

\item{112.} ``Chern-Simons term induced from a topological phase on the Wilson
fermion vacuum functional,'' Z.S. Ma, S.S. Wu, and H.Z. Li, Phys. Rev. D {\bf
52}, 337-339 (1995). (A)

\item{113.} ``Photon-echoes and Berry phase,'' R. Friedberg and S.R. Hartmann,
Phys. Rev. A {\bf 52}, 1601-1608 (1995). (I)

\item{114.} ``Gauge-invariant reference section and geometric phase,'' A.K.
Pati, J. Phys. A {\bf 28}, 2087-2094 (1995). (I)

\item{115.} ``An adiabatic phase in scattering,'' G. Ghosh, Phys. Lett. A
{\bf 210}, 40-44 (1996). (I)

\item{{\bf B.}} {\bf Experimental articles}

\item{116.} ``Manifestations of Berry's topological phase for the photon,''
R.Y. Chiao and Y.S. Wu, Phys. Rev. Lett. {\bf 57}, 933-936 (1986). (I)

\item{117.} ``Observation of Berry's topological phase by use of an optical
fiber,'' A. Tomita and R.Y. Chiao, Phys. Rev. Lett. {\bf 57}, 937-940 (1986).
(I)

\item{118.} ``Berry phase in magnetic-resonance,'' D. Suter, G.C. Chingas, R.A.
Harris, and A. Pines, Molecular Phys. {\bf 61}, 1327-1340 (1987). (I)

\item{119.} ``Adiabatic rotational splitting and Berry's phase in
nuclear-quadrupole resonance,'' R. Tycko,
Phys. Rev. Lett. {\bf 58}, 2281-2284 (1987). (I)

\item{120.} ``Manifestation of Berry topological phase in neutron spin
rotation,''
T. Bitter and D. Dubbers, Phys. Rev. Lett. {\bf 59}, 251-254 (1987). (I)

\item{121.} ``Path-dependence of the
geometric rotation of polarization in optical
fibers,'' F.D.M. Haldane, Optics letters {\bf 11}, 730-732 (1986). (E)

\item{122.} ``Measurement of the Pancharatnam phase for a light-beam,'' T.H.
Chyba, L. Mandel, L.J. Wang, and R. Simon, Optics Letts. {\bf 13}, 562-564
(1988). (E)

\item{123.} ``Observation of non-integrable geometric phase on the Poincar\'e
sphere,'' R. Bhandari, Phys. Lett. A {\bf 133}, 1-3 (1988). (I)

\item{124.} ``Observation of Berry geometrical phase in
electron-diffraction from
a screw dislocation,'' D.M. Bird and A.R. Preston,
Phys. Rev. Lett. {\bf 61}, 2863-2866 (1988). (I)

\item{125.} ``Observation of topological
phase by use of a laser interferometer,''
R. Bhandari and J. Samuel, Phys. Rev. Lett. {\bf 60}, 1211-1213 (1988). (E)

\item{126.} ``Observation of a topological phase by means of a nonplanar
Mach-Zehnder Interferometer,'' R.Y. Chiao, A. Antaramian, H. Nathel, S.R.
Wilkinson, K.M. Ganga, and H. Jiao, Phys. Rev. Lett. {\bf 60}, 1214-1217
(1988). (E)

\item{127.} ``Study of the Aharonov-Anandan Quantum phase by NMR
interferometry,'' D. Suter, A. Pines, and K.T. Mueller, Phys. Rev. Lett.
{\bf 60}, 1218-1220 (1988). (I)

\item{128.} ``Measurement of the Berry phase with polarized neutrons,'' D.
Dubbers, Physica B/C {\bf 151}, 93-95 (1988). (I)

\item{129.} ``Measurement of Berry phase
for noncyclic evolution,'' H. Weinfurter
and G. Badurek, Phys. Rev. Lett. {\bf 64}, 1318-1321 (1990). (I)

\item{130.} ``Geometric phase experiments in optics -- a unified description,''
R. Bhandari, Current Science {\bf 59}, 1159-1167 (1991). (I)

\item{131.} ``Observation of a nonclassical Berry phase for the photon,'' P.G.
Kwiat and R.Y. Chiao, Phys. Rev. Lett. {\bf 66}, 588-591 (1991). (I)

\item{132.} ``Manifestations of Berry's phase in image-bearing optical beams,''
M. Segev, R. Solomon, and A. Yariv, Phys. Rev. Lett. {\bf 69}, 590-592
(1992). (A)

\item{133.} ``A geometric-phase interferometer,'' P. Hariharan and M. Roy, J.
Mod. Optics {\bf 39}, 1811-1815 (1992). (I)

\item{134.} ``The geometric phase -- a simple optical demonstration,'' P.
Hariharan, Am. J. Phys. {\bf 61}, 591-594 (1993). (E)

\item{135.} ``The geometric phase observations at the single-photon level,'' P.
Hariharan, M. Roy, P.A. Robinson, and J.W. Obyrne, J. Mod. Optics {\bf 40},
871-877 (1993). (A)

\item{136.} ``Scheme for measuring a Berry phase in an atom interferometer,''
M. Reich, U. Sterr, and W. Ertmer, Phys. Rev. A {\bf 47}, 2518-2522
(1993). (A)

\item{137.} ``Non-cyclic geometric phases in a proposed two-photon
interferometric experiment,'' J. Christian and A. Shimony, J. Phys. A
{\bf 26}, 5551-5567 (1993). (I)

\item{138.} ``The geometric phase -- interferometric observations with
white-light,'' P. Hariharan, K.G. Larkin, and M. Roy, J. Mod. Phys. {\bf 41},
663-667 (1994). (I)

\item{139.} ``Observation of a nonlocal Pancharatnam phase-shift in the process
of induced coherence without induced emission,'' T.P. Grayson, J.R. Torgerson,
and G.A. Barbosa, Phys. Rev. A {\bf 49}, 626-628 (1994). (A)

\item{140.}``On measuring the Pancharatnam phase I. interferometry,'' A.G. Wagh
and V.C. Rakhecha, Phys. Lett. A {\bf 197}, 107-111 (1995). (I)

\item{141.}``On measuring the Pancharatnam phase II. SU(2) polarimetry,'' A.G.
Wagh and V.C. Rakhecha, Phys. Lett. A {\bf 197}, 112-115 (1995). (I)

\item{142.}``Observation of nonadiabatic
geometrical effects in a time-of-flight
experiment with polarized neutrons,'' D.A. Korneev, V.I. Bodnarchuk, and L.S.
Davtyan, Physica B {\bf 213}, 993-995 (1995). (I)

\vskip .5cm
\noindent{\it Note Added}
\vskip .5cm

After this resource letter was written the following interesting papers
on the geometric phase have been brought to our attention:

\item{143.} ``Classical Adiabatic holonomy in a Grassmannian
system,'' E. Gozzi, Phys. Rev.D. 35 (8) 2388 (1987). (A)

\item{144.}``SU(2) phase jumps and geometric phases,'' R. Bhandari, Phys.
Lett. A {\bf 157}, 221 (1991). (I)

\item{145.}``Topological Aspects of the Non-Adiabatic Berry Phase," Ali
Mostafazadeh and Arno Bohm,
J. Phys. A: Math. and Gen., {\bf 26}, 5473-5479 (1993). (A)

\item{146.}``The Relation between the Berry and the Anandan-Aharonov
 Connections for U(N) bundles," A. Bohm and A. Mostafazadeh, J. Math.
Phys., {\bf 35,} 1463-1470
(1994). (A)

\item{147.}``Geometric Phase, Bundle Classification, and Group
Representation," A. Mostafazadeh, J. Math. Phys., 37, 1218-1233 (1996).
(A)

\item{148.}``Polarization of light and topological phases,'' R.  Bhandari,
Physics Reports {\bf 281}, 1 (1997). (I)
     
\end